\documentclass [11pt,twoside,a4paper]{article}

\usepackage{shrthnds}
\usepackage{cite}
\usepackage{graphicx}
\usepackage[usenames]{color}
\usepackage{subfigure}
\usepackage{setspace}

\usepackage{multirow}

\usepackage[hang,small]{caption}

\usepackage{geometry}
\usepackage{fancyhdr}
\usepackage{titlesec,titletoc}
  \titleformat{\section}{\Large\sf\bfseries}{\thesection}{1em}{}
  \titleformat{\subsection}{\large\sf\bfseries}{\thesubsection}{1em}{}

\title{\sf\bfseries Constraints on the Cosmological Constant due to Scale Invariance}

\author{\normalsize Pavan K. Aluri$^1$\footnote{email: aluri@iitk.ac.in}~, Pankaj Jain$^{1}$\footnote{email: pkjain@iitk.ac.in}~,
 Subhadip Mitra$^{2}$\footnote{email: subhadip@imsc.res.in}~,\\ Sukanta
Panda$^3$\footnote{email: sukanta@iiserbhopal.ac.in} and Naveen K. Singh$^1$\footnote{email: naveenks@iitk.ac.in}}

\date{}

\begin{document}
 \maketitle
\vspace{-0.6cm}
\bc
{\small 1) Department of Physics, IIT Kanpur, Kanpur 208 016, India\\
2) The Institute of Mathematical Sciences, Chennai 600 113, India\\
3) Indian Institute of Science Education and Research, Bhopal 462 023, India}\ec

\centerline{\small\date{\today}}
\vspace{0.5cm}

\bc
\begin{minipage}{0.9\textwidth}\begin{spacing}{1}{\small {\bf Abstract:}
We consider the standard model with local scale invariance. The theory shows
exact scale invariance of dimensionally regulated action. We show 
that massless gauge fields, which may be abelian or non-abelian, lead to 
vanishing contribution to the cosmological constant in this
theory. This result follows in the quantum theory, to all orders in the 
gauge couplings. However we have not considered contributions 
higher orders in the 
gravitational coupling. 
Similarly we also find that massless fermion fields yield null
contribution to the cosmological constant. The effective
cosmological constant in this theory is non-zero due to the 
phenomenon of cosmological symmetry breaking, which also gives masses to
all the massive fields, besides generating the Planck mass. We find a
simple relationship between the curvature scalar and the vacuum
value of the Higgs field in the limit when we ignore all other contributions
to the energy density besides the vacuum energy.

}\end{spacing}\end{minipage}\ec

\vspace{0.5cm}\begin{spacing}{1.1}

\section{Introduction}
The idea that scale invariance may be implemented as a local symmetry was
originally suggested by Weyl \cite{Weyl:1929}. The idea was later revived
by several authors \cite{Dirac1973,Sen:1971,Utiyama:1973,Utiyama:1974,Freund:1974,Hayashi:1976,Hayashi:1978,Nishioka:1985,Ranganathan:1987}. 
A generalized standard model of elementary particles, which displays
local scale invariance, has been proposed in Ref. \cite{ChengPRL}. 
As expected, the model predicts the existence of the Weyl vector meson. 
However the Higgs particle disappears from the particle spectrum and acts
like the longitudinal model of the Weyl meson. Phenomenological consequences
of the Weyl vector meson have been studied in Refs. 
\cite{Huang1989,Wei2006,JMS,AJS}. Local scale invariance has also 
been considered in Refs. \cite{Padmanabhan85,Mannheim89,Hochberg91,Wood92,Wheeler98,Feoli98,Pawlowski99,Nishino2009,Demir2004,Mannheim09}.

However a fundamental 
problem with the model is the possibility that it may not be consistent
quantum mechanically due to scale anomaly. In Ref. \cite{Englert:1976}
it was shown that it is possible to suitably extend the scale symmetry
to $d$ dimensions such that the extended symmetry is not anomalous. 
This was also conjectured earlier in Ref. \cite{Kallosh}. This possibility
was later studied by several authors \cite{Fradkin:1978,Fradkin:1981,JMS,JM09_2,JM09_1,Shaposhnikov:2008a,Shaposhnikov:2008b}. In Ref. \cite{JM09_1}, the
authors argued that the scale invariant extension of the
 Standard model, proposed
in Ref. \cite{ChengPRL}, may be consistent
quantum mechanically. 
The authors argued that the 
model provides a very simple
and elegant solution to the standard gauge hierarchy problem. The hierarchy 
problem is solved since the model contains no physical scalar fields. The
scale invariance in this model is broken by the phenomenon of 
cosmological symmetry breaking \cite{JM,JMS,JM09_1,AJS}. Here the scale 
invariance is broken by a time dependent classical solution with non-zero
space-time curvature. This possibility has also been considered earlier by
many authors in the context of global scale invariance \cite{Shore,Cooper,Allen,Ishikawa,Buchbinder85,Wetterich1,Perez,Finelli,Odintsov:1990,Odintsov:1991,Buchbinder,Odintsov:1993,Wetterich2}. The breakdown of scale invariance in this
theory generates the gravitational constant, the cosmological constant
as well as the masses of the Standard model particles. 

The fact that scale anomaly is absent in this theory does not conflict
with the standard result that the scale symmetry is anomalous \cite{Coleman,Collins,Fujikawa}. The scale anomaly is found to be absent in this theory
only in the case when the symmetry is broken, either spontaneously or
cosmologically \cite{JM,JMS,JM09_1,AJS}. In this case the vacuum value of the scalar field or
the background curvature provides the scale needed to regulate the action
while preserving scale symmetry. However in the limit when these vacuum
values vanish, i.e. the symmetry is unbroken, the resulting
regulated action becomes ill-defined. This is the limit when the standard
result regarding the anomalous nature of the scale symmetry is applicable.

The scale invariant Standard model \cite{ChengPRL,JMS,AJS,JM09_1} may be 
consistent with cosmological observations since it predicts
dark energy and dark matter. It may provide a solution to the 
cosmological constant problem \cite{Weinberg,Peebles,Padmanabhan,Copeland,Carroll,Sahni,Ellis} due to exact scale invariance, which
does not allow a cosmological constant term to appear in the action
\cite{JMS,Wetterich,Wetterich2,JM09_2}.
Some of the parameters in the model take very small values in order 
to fit the observed dark energy, the Hubble constant, the gravitational
constant and the electroweak symmetry breaking scale. The model
provides no explanation for the smallness of these parameters. 
Despite the presence of such small parameters, the model may
not suffer from fine tuning problems if these parameters are stable
against quantum corrections \cite{JMS,JM09_2,JM09_1}. 
This might happen due to the exact scale 
invariance in this theory. In the present paper we consider the constraints
that scale invariance imposes on the effective cosmological constant
in this theory. 
Some alternate approaches to solve the cosmological
constant problem are discussed in Refs. \cite{Weinberg,Aurilia,VanDer,Henneaux,Brown,Buchmuller,Henneaux89,Sorkin,Sahni1,Sundrum}.

The action for the scale invariant extension
of the Standard Model in
$d$ dimensions may be written as \cite{JM09_1},
\ba
\mathcal{S} &=& \int d^dx \sqrt{-\bar g}\Bigg[-{\beta\over 4} \mc H^\dag \mc H
\bar R'
+\bar g^{\mu\nu} (D_\mu \mc H)^\dag(D_\nu \mc H) - \frac14
\bar g^{\mu\nu}\bar g^{\alpha\beta}(\mathcal{A}^i_{\mu\alpha} \mathcal{A}^i_{\nu\beta}
\nn\\
&+& \mathcal{B}_{\mu\alpha} \mathcal{B}_{\nu\beta} + \mathcal{G}^j_{\mu\alpha}
 \mathcal{G}^j_{\nu\beta}) (\bar R'\,^2)^{-\epsilon/4}
 - {1\over 4}\bar g^{\mu\rho}\bar g^{\nu\sigma}\mathcal{E}_{\mu\nu}\mathcal{E}_{\rho\sigma}
(\bar R'\,^2)^{-\epsilon/4} \nn\\
 &-& \lambda    (\mc H^\dag \mc H)^2 (\bar R'\,^2)^{\, \epsilon/4}  \Bigg]
 + {\cal S}_{\rm fermions},
\label{eq:S_EW_d}
\ea
where $\mc{H}$ is the Higgs doublet, $\mc{G}^j_{\mu\nu}$, $\mc{A}^i_{\mu\nu}$,
$\mc{B}_{\mu\nu}$ and $\mc{E}_{\mu\nu}$
represent the field strength tensors for the $SU(3)$, $SU(2)$, $U(1)$ and
the Weyl
vector fields respectively. The superscripts $i$ and $j$ on $\mc{A}^i_{\mu\nu}$
and $\mc{G}^j_{\mu\nu}$ represent the $SU(2)$ and $SU(3)$ indices respectively.
We are implicitly summing over these indices.
The symbol
$\bar R'$ represents the scale covariant curvature scalar,
\ba
{\bar R'}={\bar R} + 4\frac{1-d}{d-2}  S_{;\mu}^\mu + (3d-d^2-2) 
\left(\frac{2}{d-2}\right)^2f^2  S^{\mu}S_{\mu}\, 
\ea
and $f$ is the coupling of the Weyl vector bosons.
Here we use the notation of Ref. \cite{tHooft} and denote all quantum gravity
variables with a bar. In Eq. \ref{eq:S_EW_d}, $\beta$ and $\lambda$ denote the
coupling parameters.
The regulated fermionic action in d-dimensions is given by,
\ba
{\cal S}_{\rm fermions} &=&\int d^d x\, e\, \left({\overline\Psi}_{\rm L}i
\gamma^\mu  {\cal D_\mu} \Psi_{\rm L} +
{\overline\Psi}_{\rm R}i \gamma^\mu  {\cal D_\mu} \Psi_{\rm R}  \right)\nn\\
&&- \int d^dx\, e\, (g_Y\overline{\Psi}_{\rm L} {\mc H}\Psi_{\rm R} (\bar R'\,^2)^{\, \epsilon/8} + h.c.),
\ea
where $e={\rm det}(e_\mu^{~a})$, $e_\mu^{~a}e_\nu^{~b}\eta_{ab}
=g_{\mu\nu}$, $\gamma^\mu=e^\mu_{~a}\gamma^a$  and
$a, b$ are Lorentz indices.
Here $\Psi_{\rm L}$ is an $SU(2)$ doublet, $\Psi_{\rm R}$ a singlet and $g_Y$ represents
a Yukawa coupling. For simplicity we have displayed only one Yukawa
coupling term. Furthermore we have displayed the action only
for a single family.

The covariant derivative acting on the fermion field is defined by
\begin{equation}
{\cal D}_\mu \Psi_{\rm L,R} = \left({\tilde D}_\mu + {1\over2}\omega_{\mu}^{ab}\sigma_{ab}\right)\Psi_{\rm L,R}\ ,
\label{eq:Dfermion}
\end{equation}
where ${\tilde D}_{\mu}\Psi_{\rm L} = \partial_\mu -i g{\bf T\cdot A}_\mu - ig'{Y^{\rm L}_{f}\over 2} B_\mu$,
${\tilde D}_{\mu}\Psi_{\rm R} = \partial_\mu - ig'{Y^{\rm R}_{f}\over 2} B_\mu$
and $\sigma_{ab} = {1\over4}[\gamma_a,\gamma_b].$ In Eq. \ref{eq:Dfermion},
$A_\mu$ is the $SU(2)$ field, $B_\mu$ the $U(1)$ field, ${\bf T}$ represents
the SU(2) generators and $Y_f$'s the U(1) hypercharges. Here we have not
explicitly displayed the color interactions for quarks, which can be easily
added. The spin connection $\omega_{\mu}^{ab}$ can be solved in terms
of the vierbein,
\ba
\omega_{\mu ab}\ =\ {1\over 2}(\partial_\mu e_{b\nu}-\partial_\nu e_{b\mu})e_a^{~\nu}
-{1\over2}(\partial_\mu e_{a\nu}-\partial_\nu e_{a\mu})e_b^{~\nu}
-{1\over2}e_a^{~\rho}e_b^{~\sigma}
(\partial_\rho e_{c\sigma}-\partial_\sigma e_{c\rho})e^c_{~\mu}\ .
\ea
The spin connection term in covariant derivative makes the fermionic action
locally scale invariant. The
Weyl vector boson does not couple directly to fermions. However it couples
indirectly since the factor $\bar{R}'$ contains contribution from the Weyl
meson.

In this paper we shall obtain constraints on the cosmological constant
that are imposed due to scale invariance. The model does not permit a 
cosmological constant term in the action. We expect that classically
the trace of the energy momentum tensor would vanish due to conservation
of the scale current $J^\mu$, i.e.
\ba
(J^\mu)_{;\mu} = T^\mu_\mu = 0,
\ea
where $T_{\mu\nu}$ is the energy momentum tensor. This would normally lead to
direct constraints on the energy density of the matter fields. In particular
this would imply that the
vacuum energy density is zero. However due to non-minimal coupling of the
matter fields to gravity this
does not necessarily follow. The scale invariance is broken by the phenomenon
which we refer as the cosmological symmetry breaking \cite{JM,JMS}. 
This leads to a non-zero value for the cosmological constant. Since the model
does not permit a cosmological constant term, the value of this parameter
generated due to symmetry breaking is a prediction of the model. 

In the present paper we shall determine the constraints that scale invariance
imposes on the cosmological constant in the full quantum theory. We shall
work within the framework of canonical quantization. The fact that we
can regulate the action while preserving scale invariance implies that
scale symmetry is exact in the full quantum theory. 
A basic observable of cosmological interest is the curvature scalar $R$. 
The curvature
scalar is related to the energy momentum
tensor by the Einstein's equations. As we shall see, in the present theory, 
it is also related to the matter fields by the 
Higgs field equation of motion. 
We point out that this equation is valid in the full
quantum theory as the Heisenberg operator equation \cite{Nishijima}. All operators arising 
in this equation are well defined since the equation is obtained from  
the regulated action. It turns out that this equation provides us with a 
much simpler equation for determining $R$.

In our analysis we shall focus primarily on the contributions due to the
Standard Model fields. We shall not explicitly include the contributions
due to the Weyl vector meson. These depend on another unknown parameter
$f$. The contribution of Weyl vector meson to the cosmological constant
will be small as long as this parameter is sufficiently small. However we shall
not consider the precise observational constraint on this parameter in
this paper. Furthermore we shall also ignore quantum gravity effects.

In the next section we consider a simple system of real scalar field
coupled to gravity. In section 3 we analyse the energy momentum
tensor in the Standard Model.

\section{A Toy Model}
In order to get oriented we first consider a simple system of a real scalar
field coupled to gravity. The action may be written as
\begin{equation}
\mc{S} = \int d^4x \sqrt{-g}\left[\frac12g^{\mu\nu}\partial_\mu\Phi\partial_\nu\Phi
-\frac{\lambda}{4}\Phi^4 -\frac{\beta}{8}\Phi^2R \right],
\label{eq:Action_Classical}
\end{equation}
The Einstein's equations are
\begin{equation}
\Phi^2\left[-{1\over 2} g_{\alpha\beta} R + R_{\alpha\beta}\right]
+(\Phi^2)_{;\lambda;\kappa}\left[-{1\over 2} (g_\alpha^\lambda g_\beta^\kappa
+ g_\alpha^\kappa g_\beta^\lambda) + g_{\alpha\beta} g^{\lambda\kappa}\right]
= {4\over \beta} T_{\alpha\beta},
\label{eq:Einst_eom1}
\end{equation}
where
\be
T_{\al\bt} = \pr_\al\Ph\pr_\bt\Ph - g_{\al\bt}\left[
{1\over 2} g^{\m\n} \pr_\m\Ph\pr_\n\Ph-{\frac\lm4} \Phi^4\right].\label{eq:std_emtensor}
\ee
The equation of motion for the scalar field is given by,
\begin{equation}
g^{\mu\nu}\Phi_{;\mu;\nu} + {\beta\over 4}\Phi R + \lambda\Phi^3 = 0.
\label{Eq:scalar_eom}
\end{equation}
We point out that $T_{\alpha\beta}$ is not the total energy momentum tensor
since it does not include the contribution from the term proportional to
$\beta$. We shall refer to $T_{\alpha\beta}$ as the truncated energy
momentum tensor. In this theory the scalar field is not separable from the
gravitational action. Hence the total energy momentum tensor would
necessarily involve contributions from the gravitational sector also.
We may determine the energy momentum tensor by varying the action with
respect to the gravitational field and identifying the terms linear
in the fluctuations $\delta g_{\mu\nu}$. The corresponding tensor, and hence
its trace, vanishes trivially by the Einstein's equations.

The action given in Eq. \ref{eq:Action_Classical} displays the phenomenon of
cosmological symmetry breaking. We assume a FRW metric with the curvature
parameter $k$ set to zero, for simplicity. We find a classical solution
with the scalar field $\Phi_0$ equal to a constant and
\ba
{\beta\over 4}R = -\lambda \Phi_0^2.
\label{eq:classical_Phi}
\ea
 The FRW scale parameter is found to be,
\ba a(t) = a_0 \exp(H_0t)
\label{soln_a(t)}
\ea
with
\ba \Phi_0 = \sqrt{3\beta\over\lambda} H_0,
\label{soln_Phi}
\ea
where $H_0$ is identified with the Hubble constant. This is a time dependent
solution to the equations of motion. If we assume the Big Bang model for
the universe then all physical phenomena take place in this background.
Here we have generated an effective cosmological constant equal to 
$\lambda \Phi_0^4/4$ due to the non-minimal coupling of the scalar
field with gravity. Specifically we mean the term proportional to $\Phi^2R$
in the action. 
In order to study any process we need to make a quantum expansion
around the classical background $\Phi_0$. 
We point out that since we are not making
an expansion around the minimum of the potential, the lowest energy state
is not the true ground state of the theory.
In order to obtain the curvature scalar in the full quantum theory, we need
to take the expectation value of Eq. \ref{eq:Einst_eom1} with respect to
the lowest energy state. 
We shall consider the adiabatic
limit where the time dependence of the classical solution is very slow.
Hence we shall drop terms which are higher order in the
derivative of the metric.
These will introduce additional powers of the Hubble constant $H_0$ which
is the smallest energy scale in the theory. In fact all other energy scales
are related to this fundamental scale but these will involve factors of
$1/\lambda$ or $\beta$ which we assume to be large. We shall also ignore
higher order corrections in quantum gravity.

We next consider the Einstein's equation for the toy model. 
It is useful to define the tensor
$\tilde T_{\mu\nu}$, which collects all the terms in the Einstein's equation, 
besides the term proportional to the Einstein
tensor $(R_{\mu\nu} - g_{\mu\nu} R/2)$. We have
\begin{equation}
\tilde T_{\alpha\beta} = T_{\alpha\beta} - {\beta\over 4}
(\Phi^2)_{;\lambda;\kappa}\left[-{1\over 2} (g_\alpha^\lambda g_\beta^\kappa
+ g_\alpha^\kappa g_\beta^\lambda) + g_{\alpha\beta} g^{\lambda\kappa}\right].\label{eq:tildeT}
\end{equation}
We are interested in evaluating trace of this tensor. 

We expand the full quantum metric,
denoted by $\bar g_{\mu\nu}$, around the classical solution, $g_{\mu\nu}$,
\ba
\bar{g}_{\mu\nu} = g_{\mu\nu} + h_{\mu\nu}.
\label{eq:metric_ex}
\ea
Here we treat gravity classically and only the matter fields are quantized.
Hence in the expansion of the action we shall only keep terms linear in
the field $h_{\mu\nu}$. We may express the terms linear in $h_{\mu\nu}$ as,
\begin{equation}
\delta S = {1\over 2} \int d^4 x \sqrt{-g} \left[{\beta\over 4}\Phi^2\left(R_{\mu\nu}-
g_{\mu\nu} R/2 \right) - {\tilde T}_{\mu\nu}\right] h^{\mu\nu}
\label{eq:tilde_T}
\end{equation}
The Einstein's equations are obtained by demanding that the terms inside
the square brackets vanish.
We expand the field $\Phi$ around the classical solution $\Phi_0$,
\begin{equation}
\Phi = \Phi_0 + \phi\,.
\end{equation}
We point out that the classical solution $\Phi_0$ is constant.
Since we are interested in a quantum calculation 
we must suitably regulate the action. We use
dimensional regularization, which introduces an extra factor $(R^2)^{\epsilon/4}$ in self coupling term proportional to $\lambda$, in analogy with the
regulated action shown in Eq. \ref{eq:S_EW_d}.

Before we proceed we note that the term proportional to $(\Phi^2)_{;\lambda;\kappa}$ in Eq. \ref{eq:tildeT} does not contribute to the cosmological constant.
The reason is the following - the contribution of this term to the action is proportional to
\bas\int d^dx\sqrt{-g} \left[(\Phi^2)_{;\lambda}\right]^{;\lambda} h
= \int d^dx \left[\sqrt{-g}(\Phi^2)_{;\lambda}\right]^{,\lambda} h\,, \eas
where $h=h_\mu^\mu$.
After integrating by parts this term becomes proportional to
a derivative of $h$ which means that it can contribute only if the external
graviton has a non-zero momentum.
Hence it contributes zero to the cosmological
constant. Same argument applies to all 
terms in the Lagrangian density of the form 
$\sqrt{-g} (V_\alpha)_{;\beta} h^{\alpha\beta}$, where $V_\alpha$ is a 
vector. A term in the Lagrangian density of the form $\sqrt{-g} (V_\alpha)_{;\beta} 
g^{\alpha\beta}$, where $g^{\alpha\beta}$ is the classical metric, vanishes trivially by integration by parts. 
Hence we find the general result that all terms in the energy momentum tensor
of the form $(V_\alpha)_{;\beta}$ give zero contribution to the cosmological
constant.

We next consider the trace of $T_{\alpha\beta}$. We may express the trace as
\begin{equation} 
T_\alpha^\alpha = \left(1-{\epsilon\over 2}\right)\left[-
(\Phi \Phi^{;\mu})_{;\mu} + \Phi \Phi_{;\mu}^{;\mu}
+\lambda\Phi^4 (R^2)^{\epsilon/4}\right] + \ldots\,.
\label{eq:Talphaalpha1}
\end{equation}
Here we have not explicitly displayed some terms which are proportional
to an overall derivative and hence do not contribute to cosmological constant.
The first term on the right side would also not contribute to the cosmological
constant for the same reason.
Taking the trace of the Einsteins's equation, Eq. \ref{eq:Einst_eom1}, 
in $d$ dimensions, we find, 
\begin{equation}
g^{\mu\nu}\Phi\Phi_{;\mu;\nu} + {\beta\over 4}\Phi^2 R + \lambda\Phi^4(R^2)^{\epsilon/4} + ... = 0.
\label{eq:toy_R}
\end{equation}
where we have used Eq. \ref{eq:Talphaalpha1}. Here, once again, we have dropped the surface
terms which do not contribute to the cosmological constant.
The result of Eq. \ref{eq:toy_R}, actually follows much more directly from the equation of motion of the scalar
field, i.e., Eq. \ref{Eq:scalar_eom} suitably generalized to $d$-dimensions. This is a consequence of the non-minimal 
coupling of the scalar field to gravity.

Classically the curvature scalar $R$ is determined by Eq. 
\ref{eq:classical_Phi}. In the quantum theory we need to compute the
higher order corrections to this result. 
We may choose the renormalization scheme such that the expectation
value of $\Phi$ in the lowest energy state,
\ba
<\Phi> = \Phi_0
\label{eq:expect_Phi}
\ea
and $<\phi> = 0$ to all orders.
This implies that
\ba
<\Phi>^2 = -{\beta\over 4\lambda} R\,,
\label{eq:expect_Phi2}
\ea
to all orders.
This equation directly fixes the value of the curvature $R$ in terms of the
renormalized parameters $\beta$, $\lambda$ and $<\Phi>$.
It is valid as long as we include contributions only due to vacuum energy.

We may impose local scale invariance on this toy model by introducing the Weyl vector meson $S_\mu$. In this case
the real scalar field disappears from the
particle spectrum and acts like the longitudinal mode of the Weyl meson.
We again expand around a classical solution
$\Phi = \Phi_0+\phi$, with $\Phi_0$ equal to a constant. 
Furthermore lets assume that $<S_\mu>=0$, i.e. the expectation value of the Weyl meson field in the lowest
energy state is exactly equal to zero. 
As we expand around the classical solution, the one
point function $<\phi>$ has to vanish in this case since the scalar particle
is unphysical. Hence we directly obtain Eq. \ref{eq:expect_Phi2} at all
orders, which fixes $R$ in terms of other renormalized parameters. We
study the implications of this in the next section where we study
the model defined by  Eq. \ref{eq:S_EW_d}.

\section{Constraints on the Cosmological Constant}
We now obtain the constraints on the cosmological constant due to
scale invariance for the standard model defined by Eq.
\ref{eq:S_EW_d}. Here we shall treat gravity classically.
The Einstein's equation may be written as,
\ba
\mc H^\dag \mc H \left[-{1\over 2} { g}_{\alpha\beta} R + {R}_{\alpha\beta}\right]
+(\mc H^\dag \mc H)_{;\lambda;\kappa}\left[-{1\over 2} ({g}_\alpha^\lambda { g}_\beta^\kappa
+ {g}_\alpha^\kappa {g}_\beta^\lambda)
+ { g}_{\alpha\beta} {g}^{\lambda\kappa}\right]
= {2\over \beta} T_{\alpha\beta},
\label{eq:Einst_eom2}
\ea
where the truncated energy momentum tensor, $T_{\alpha\beta}$, 
contains contributions due to all the terms excluding
the term proportional to $\beta$ in the action. 
We are interested in extracting 
the effective cosmological constant or equivalently the curvature scalar $R$
from this equation. 
The contribution of the second term on the left hand side of 
Eq. \ref{eq:Einst_eom2} to $R$ vanishes in analogy to the corresponding term 
in the toy model considered in 
Section 2. 

\subsection{Pure Higgs Field}
Let us first consider the scalar Higgs field coupled to gravity. We ignore all
other fields for now. This is very similar to the toy model considered
in last section, except that we also take into account the regulator and
we are dealing with a complex doublet.
If we include contribution only from the Higgs field we find,
\ba
T_{\alpha\beta} &=& (D_\alpha \mc H)^\dag D_\beta\mc H + (D_\bt \mc H)^\dag D_\al\mc H - g_{\alpha\beta}
\left[ g^{\mu\nu} (D_\mu \mc H)^\dag D_\nu\mc H - \lambda
(\mc H^\dag \mc H)^2 (R^2)^{\epsilon/4}\right] \nn\\
&&-\ep\lm(\mc H^\dag \mc H)^2(R^2)^{(\ep-2)/4}R_{\al\bt}+ \ldots\,.
\label{eq:scalarEMT}
\ea
Here the `\ldots' denotes terms which are proportional to an overall
derivative besides terms proportional to the Weyl meson coupling $f$.
As mentioned in the introduction we shall not explicitly consider the contribution due to
Weyl meson in this paper. The terms in $T_{\alpha\beta}$  which are of the 
form $(V_\alpha)_{;\beta}$, where $V_\alpha$ is a vector, do not contribute to the cosmological constant. 
These terms are proportional to an overall derivative and as shown in Section 2, their contributions vanish.

The trace of the remaining terms, explicitly displayed in Eq. \ref{eq:scalarEMT}, may be expressed as
\ba
T^\alpha_\alpha &=&\left(1- {\epsilon\over 2}\right)\left[ -2D_\mu(g^{\mu\nu} \mc H^\dag D_\nu \mc H) + 2g^{\mu\nu}
\mc H^\dagger D_\mu D_\nu \mc H + 4\lambda(\mc H^\dag \mc H)^2 (R^2)^{\epsilon/4}\right]
+ \ldots\,.
\ea
The first term on the right hand side is a total derivative and can be
ignored. If we take the trace of the Einstein's equation, Eq. 
\ref{eq:Einst_eom2}, and substitute $T_\alpha^\alpha$ into the resulting
equation we obtain,
\ba
\mc H^\dagger D^\mu D_\mu \mc H + {\beta\over 4}R\mc H^\dagger \mc H + 
2\lambda (\mc H^\dag\mc H)^2 
(R^2)^{\epsilon/4} +... = 0\,.
\ea
As before, this result can be obtained much more directly by using the 
equation of motion of the Higgs field coupled to gravity, i.e.,
\ba
D^\mu D_\mu \mc H + {\beta\over 4}R\mc H + 2\lambda  \mc H(\mc H^\dag\mc H)
(R^2)^{\epsilon/4} = 0\,.
\ea

\subsection{Pure Gauge Fields}
Next, we consider the contribution due to a pure non-abelian  
gauge field (the calculation for an abelian gauge field is similar except the fact that there is no Faddeev-Popov ghost term).
The canonical quantization of non-abelian gauge fields is discussed in
Refs. \cite{Nakanishi,Haller,ChengTsai}.
The corresponding action
may be written as,
\ba
\mathcal{S}_{\rm A} = \int d^dx \sqrt{- g}\left[
 - \frac14  g^{\mu\nu} g^{\alpha\beta}(\mathcal{A}^i_{\mu\alpha}
 \mathcal{A}^i_{\nu\beta})( R^{\,2})^{-\epsilon/4}+{\mathcal L}^{gf} + 
 {\mathcal L}^{\rm FPG}\right] ,
\label{eq:S_gaugeKE}
\ea
where $\mathcal{A}^i_{\mu\nu}$ represents the field strength tensor of the gauge
field and the superscript $i$ represents the internal index. Summation over 
$i$ is implicit. The gauge fixing and the Faddeev-Popov ghost Lagrangian
are given by,
\begin{equation}
\mathcal{L}^{gf} = -{1\over 2\xi} g^{\mu\nu}g^{\alpha\beta} D_\mu A_\nu^a
D_\alpha A_\beta^a (R^2)^{-\epsilon/4}
\end{equation}
and
\begin{equation}
\mathcal{L}^{\rm FPG} = c_a^\dagger g^{\alpha\beta} D_\alpha \left[\delta_{ab}
D_\beta - gC_{abc} A_\beta^c\right]c_b
\end{equation}
respectively. Here $\xi$ is the gauge parameter, $c_a$, $c_a^\dagger$ the
ghost fields, $D_\mu$ the covariant derivative, $g$ the gauge coupling 
and $C_{abc}$ represents the gauge group structure constants.  

The contribution of the kinetic energy term to the energy momentum
tensor is given by
\ba
T^{\rm A}_{\sigma\rho} &=& {1\over 4} g_{\sigma\rho}
[g^{\mu\nu}g^{\alpha\beta} \mc {A}^i_{\mu\alpha}\mc {A}^i_{\nu\beta}
(R^2)^{-\epsilon/4}]
-  [g^{\alpha\beta} \mc {A}^i_{\sigma\alpha}\mc {A}^i_{\rho\beta} (R^2)^{-\epsilon/4}]\nonumber\\
&+&{\epsilon\over 4}[g^{\mu\nu}g^{\alpha\beta} \mc {A}^i_{\mu\alpha}\mc {A}^i_{\nu\beta} (R^2)^{-(\epsilon+2)/4}] R_{\sigma\rho}\nonumber\\ &+&
{\epsilon\over 4} [g^{\mu\nu}g^{\alpha\beta} \mc {A}^i_{\mu\alpha}\mc {A}^i_{\nu\beta} (R^2)^{-(\epsilon+2)/4}]_{;\gamma;\delta}\left[-{1\over 2}(g^\gamma_\sigma
g^\delta_\rho + g^\delta_\sigma g^\gamma_\rho) + g_{\sigma\rho}
g^{\gamma\delta}\right].
\label{eq:T2}
\ea
Once again we have ignored terms proportional to the Weyl meson coupling $f$.
In order to determine the contribution to the cosmological constant we consider
the trace of the tensor $T^{\rm A}_{\sigma\rho}$. We find that
when we take the trace of the first three terms on the right hand side
cancel identically in $d$ dimensions. This leaves us with the last, the fourth
term. 
This term of course vanishes classically when we take the limit
$\epsilon\rar 0$.
However it may contribute at loop orders.

The trace of the fourth term on the
right hand side of Eq. \ref{eq:T2} may be written as, 
\ba
{\epsilon\over 4} [g^{\mu\nu}g^{\alpha\beta} \mc {A}^i_{\mu\alpha}\mc {A}^i_{\nu\beta} (R^2)^{-(\epsilon+2)/4}]_{;\gamma}^{;\gamma}(d-1).
\ea
This term is of the form $(V^\mu)_{;\mu}$ and hence does not contribute to
the cosmological constant. 
Hence we find that the kinetic energy terms for the vector fields
do not contribute to the cosmological constant.

The contribution due to the gauge fixing term may be written as,
\ba
T^{gf}_{\alpha\beta} &=& {1\over 2\xi}\left[g_{\alpha\beta}(D\cdot A^a)^2 - 4(D\cdot A^a)(D_\alpha A_\beta^a) + \epsilon (D\cdot A^a)^2 
{R_{\alpha\beta}\over R}\right](R^2)^{-\epsilon/4}\cr
&-&{\epsilon\over 2\xi} g_{\alpha\beta} D^\mu D_\mu\left[(D\cdot A^a)^2 
{(R^2)^{-\epsilon/4}\over R}\right] + {\epsilon\over 2\xi} D_\alpha D_\beta
\left[(D\cdot A^a)^2{(R^2)^{-\epsilon/4}\over R}\right]\cr
&+&{2\over \xi} D_\alpha [A^a_\beta (D\cdot A^a) (R^2)^{-\epsilon/4}]
- {1\over \xi} g_{\alpha\beta} D^\nu[A^a_\nu (D\cdot A^a) (R^2)^{-\epsilon/4}],
\ea
where $D\cdot A^a\equiv D^\mu A^a_\mu$. When we take the trace of this 
expression we find that all the terms either cancel out or are equal to
total derivatives. After integration by parts such total derivative terms
do not contribute to the cosmological constant. Hence the contribution
of the gauge fixing term to cosmological constant also vanishes. 

Finally we consider the contribution due to the Faddeev-Popov ghost term. 
Its contribution to the energy momentum tensor may be expressed as
\ba
T^{\rm FPG}_{\mu\nu} = -g_{\mu\nu} c_a^\dagger D^\alpha \left[\delta_{ab}
D_\alpha - gC_{abc} A_\alpha^c\right]c_b + 2 c_a^\dagger D_\mu
\left[\delta_{ab} D_\nu - gC_{abc} A_\nu^c\right]c_b.
\ea
The trace of this term vanishes after using the ghost field equations of
motion. 

Hence we find that pure gauge fields yield null contribution to the 
cosmological constant.
We note the crucial use of the scale invariant dimensional regulator
used in arriving at this result. This implies that all massless gauge
fields, abelian or non-abelian, do not contribute to the trace of
the energy momentum tensor.
The result applies directly to QED and QCD gauge fields and 
is exact at all orders in the gauge coupling. However we have
not considered terms higher order in the gravitational coupling and the
Weyl meson coupling $f$.

\subsection{Massless Dirac Fermions}
The action for massless Dirac fermions in $d$ dimensions may be written as
\ba
{\cal S}_\Psi\ =\int d^dx e\, ({\overline\Psi}i\gamma^\mu  {\cal D_\mu} \Psi)
\,.
\ea
Here we ignore all the gauge fields and the covariant derivative
includes only the contribution due to gravity.  The energy momentum tensor is
found to be,
\ba
T_{\alpha\beta}^{\Psi} &=& -\,g_{\alpha\beta}
(\overline{\Psi}i\gamma^\rho {\cal D}_\rho \Psi)
+{1\over2}\overline{\Psi}i\gamma_\alpha {\cal D}_\beta\Psi
+{1\over2}\overline{\Psi}i\gamma_\beta {\cal D}_\alpha\Psi
\nonumber\\
&& + {g_{\alpha\beta}\over2}(\overline{\Psi}i\gamma_\rho\Psi)^{;\rho}
-{1\over4}(\overline{\psi}i\gamma_\beta\psi)_{;\alpha}
-{1\over4}(\overline{\psi}i\gamma_\alpha\psi)_{;\beta}.
\ea
The terms in the second line, which represent total covariant derivatives,
do not contribute to the cosmological constant. The terms in the first
line are found to be zero by using the equation of motion
\begin{equation}
i\gamma^\mu {\cal D}_\mu \Psi = 0.
\end{equation}

The above result also applies directly if we include gauge interactions
of fermions. The only modification here is that the covariant derivative
also includes the contribution from the corresponding gauge fields.
Hence we find that we get zero contribution to the cosmological constant
from both the strong and
electromagnetic interactions as long as we do not include the mass terms
for the fermions or equivalently the Yukawa interactions terms.
We include these terms below along with the electroweak gauge fields.

\subsection{The Scale Invariant Standard Model}
We now consider the  full action given in Eq. \ref{eq:S_EW_d}.
The scalar field equation of motion is given by,
\ba
D_\mu D^\mu\mc H + 2\lambda (\mc H^\dag\mc H)\mc H (R^2)^{\epsilon/4} + g_Y \bar \Psi_{\rm R}
\Psi_{\rm L} (R^2)^{\epsilon/8} +{\beta\over 4} \mc H R = 0\,.
\label{eq:scalareom}
\ea
Here we have not explicitly displayed the contribution due to the gauge
fixing terms and the Faddeev-Popov ghost terms.
The equations of motion for the $SU(2)$ doublet and singlet fermions may be written as,
\ba
i\gamma^\mu D_\mu\Psi_{\rm R} - g_Y\mc H^{\dag}\Psi_{\rm L} (R^2)^{\epsilon/8} &=& 0,\\
i\gamma^\mu D_\mu\Psi_{\rm L} - g_Y\mc H \Psi_{\rm R} (R^2)^{\epsilon/8} &=& 0.
\ea
The truncated energy momentum tensor $ T_{\alpha\beta}$ is found to be
\ba
T_{\alpha\beta} &=& -g_{\alpha\beta} \Bigg[(D_\mu\mc H)^\dag D^\mu\mc H 
+ \bar\Psi_{\rm L} i\gamma^\rho D_\rho \Psi_{\rm L}
+ \bar\Psi_{\rm R} i\gamma^\rho D_\rho \Psi_{\rm R}
- \lm (\mc H^\dag\mc H)^2 (R^2)^{\epsilon/4}\nonumber\\ 
&&- g_Y\big(\bar\Psi_{\rm L}\mc H\Psi_{\rm R}
+h.c.\big)(R^2)^{\epsilon/8}\Bigg]
+ (D_\alpha\mc H)^\dag D_\beta\mc H + (D_\bt\mc H)^\dag D_\al\mc H\nonumber\\
&&+ \frac12\bar\Psi_{\rm L} i\gamma_\alpha D_\beta\Psi_{\rm L} 
+ \frac12\bar\Psi_{\rm R} i\gamma_\alpha D_\beta\Psi_{\rm R}
+ \frac12\bar\Psi_{\rm L} i\gamma_\bt D_\al\Psi_{\rm L}
+ \frac12\bar\Psi_{\rm R} i\gamma_\bt D_\al\Psi_{\rm R}\nn\\
&&-\lm\epsilon(\mc H^\dag \mc H)^2  (R^2)^{(\epsilon -2)/4} R_{\alpha\beta}
- {\epsilon g_Y\over 2}
\big(\bar\Psi_{\rm L}\mc H\Psi_{\rm R}
+h.c.\big)  (R^2)^{(\epsilon-4)/8} R_{\alpha\beta} + \ldots,
\ea
where we have not explicitly displayed terms whose contribution to the
cosmological constant vanishes. These include the kinetic energy terms for 
the gauge fields. We have also not displayed the contributions due to the
gauge fixing terms and the Faddeev-Popov ghost terms.  
The trace of the truncated energy momentum tensor is found to be,
\ba
T^\alpha_\alpha = -(1-\epsilon/2) \left[2(D_\mu\mc H)^\dag D^\mu\mc H -
g_Y (\bar\Psi_{\rm L}\mc H\Psi_{\rm R}
+h.c.\big)  (R^2)^{\epsilon/8 }
- 4\lambda
(\mc H^\dag\mc H)^2 (R^2)^{\epsilon/4}\right] + \ldots,
\ea
where we have used the fermion equations of motion to eliminate the fermion
kinetic energy terms. 
We now substitute $T^\alpha_\alpha$ into the trace of the Einstein's equation,
Eq. \ref{eq:Einst_eom2}. 
We obtain a relatively simple expression for the trace of
the truncated energy momentum tensor or equivalently the curvature scalar. Many
of the terms cancel exactly, as expected from scale invariance. In fact
the non-zero contribution arises primarily due to the non-minimal coupling
of the Higgs field to gravity. 

In this model the cosmological constant is generated by the phenomenon of cosmological symmetry breaking. Again we assume
a FRW metric with curvature parameter $k=0$ and we expand the scalar field around its classical solution, $\mc H = \mc H_0 + \mc H'$, with,
\ba
\mc H^{\dag}_0\mc H_0 = -\frac{\beta R}{4\lambda}.
\ea
We assume that $<S_\mu> = 0$, i.e. the expectation value of the Weyl meson 
field in the lowest energy state is identically equal to zero. Following the argument given at the end of Section 2, the one point function, $<\mc H'>$, must vanish in this theory. 
This follows by gauge invariance. Since the Higgs field is
the longitudinal mode of the Weyl meson, its one point function 
must vanish as long as $<S_\mu> = 0$.  
This implies that
\ba
<\mc H> = \mc H_0 = \left(\begin{array}{c}
0\\ v
\end{array}\right)
\label{eq:VacH}
\ea
exactly to all order in the perturbation theory. Here 
\ba
v = \sqrt{-\frac{\beta R}{4\lambda}}\,.
\label{eq:v}
\ea
The parameter $\mc H_0$ in Eq. \ref{eq:VacH} is fixed by its relationship
to the $W$ boson mass,
\ba
M_{\rm W}^2 = g^2(\mc H_0^\dag\mc H_0),
\ea
where $g$ is the gauge coupling. 
In Eq. \ref{eq:VacH} we have used the electroweak symmetry to set the first entry of the classical scalar doublet to be zero. This is analogous to the
choice normally made in analysing the electroweak symmetry breaking in the standard Weinberg-Salam model.
We, therefore, find that the curvature scalar and hence the effective 
cosmological constant is determined by Eqs. \ref{eq:VacH} and \ref{eq:v} in 
the standard model with local scale invariance. 
In  Eq. \ref{eq:v}, $\beta$ and $\lambda$
are the renormalized parameters. 
The parameter $1/\beta$ is effectively the gravitational
coupling. Classically it is related to the Planck mass by the formula,
\ba
\beta\mc H^\dag_0\mc H_0 = {M_{\rm PL}^2\over 4\pi}\ .
\ea
Hence this parameter is fixed by the value of the known gravitatonal coupling.

The coupling parameter $\lambda$ does
not directly relate to any scattering cross section or the mass of any
particle since the Higgs meson is not a physical particle in this theory.  
It contributes directly to the value of the curvature scalar through 
Eqs. \ref{eq:VacH} and \ref{eq:v}. We may
fix this parameter by the observed value of the cosmological constant or
equivalently the value of the curvature scalar $R$.
Hence we find that there is sufficient freedom in the model in order to fit
the observed value of the effective cosmological constant, despite the fact
that we are not allowed to add a cosmological constant term to the action.

The model solves the gauge hierarchy problem since it does not contain
any fundamental scalar field in the physical spectrum \cite{JM09_1}.
However the model introduces three parameters, $1/\beta$, $\lambda$ and $f$, which
take very small values. The model offers no insight into why these are
driven to such small values.
Furthermore it is possible that the model may suffer from fine tuning 
problems at higher orders in perturbation theory due to the small values of
these parameters.
The parameter $1/\beta$ is equivalent to the
gravitational coupling and its smallness is well known. 
The detailed renormalization of this parameter is beyond the
scope of the present paper. Even at one loop we do not expect the 
gravitational sector of the theory to be renormalizable. 
Hence we either need to treat gravity classically or perform quantum
gravity calculations treating it as an effective theory, by introducing
an infinite number of counter terms \cite{Donoghue}. Nevertheless we do not expect
any fine tuning in the gravitational coupling. 
As we shall see
below the remaining two parameters are also intrinsically tied to the 
gravitational sector of the theory.

The renormalization of the self coupling $\lambda$ is constrained by 
the requirement that the one point function of the Higgs field must vanish 
as long as the expectation value of the Weyl meson field, $<S_\mu>$,
is zero. This requirement, imposed by gauge
invariance, relates the counterterm corresponding to the self coupling to
the gravitational counter term action.  
Hence the renormalization of $\lambda$ is intrisically tied to the 
renormalization of the gravitational field. In the absence of a consistent
theory of quantum gravity we may simply ignore higher order corrections
in $1/\beta$. In this case the theory is renormalizable and we can 
consistently study the renormalization group evolution of $\lambda$.
We point out that 
we do not expect any fine tuning in this parameter at loop orders
since it is expected to depend logarithmically on the renormalization scale.
A small value of $\lambda$ 
 may be natural due to the triviality of the pure scalar field theory. If we ignore all other fields then this parameter must vanish. However it is possible that quantum gravity effects might lead to a small
value of this parameter since they impose an effective ultraviolet cutoff on
the theory. 

Finally we consider the Weyl meson coupling $f$, which we have also assumed
to be very small. This parameter can be fixed by considering the scattering
of Weyl mesons and gravitons. We cannot consider its scattering with Standard
model particles, since all these couplings vanish. Hence we find 
that the renormalization of the coupling $f$ is also tied primarily with the
gravitational field. We expect that loop corrections to $f$ are likely to be
highly suppressed since these will involve additional powers of $f$ and/or 
the gravitational coupling.

\section{Conclusions}
In this paper we have considered an extension of the standard model of
particle physics with local scale invariance.
We have shown that scale invariance imposes considerable
constraints on the effective cosmological constant or the trace of the
energy momentum tensor. The contribution of massless gauge fields and
fermions to the cosmological constant is shown to be
identically zero. This implies that QCD (or QED) in the limit of zero fermion
masses
would yield a vanishing contribution to the cosmological constant. 
 The result follows exactly to all orders in gauge coupling. However
we have not considered higher orders in gravitational coupling or the Weyl 
coupling $f$. In hindsight the vanishing of the cosmological constant 
for massless gauge fields and fermions is not very surprising. 
It follows directly from the scalar field equation of motion, 
Eq. \ref{eq:scalareom}, which 
provides us with another equation, besides the Einstein's equations,
to compute the curvature scalar $R$. The massless gauge fields or fermions
do not contribute to this equation and hence do not contribute to $R$ and 
equivalently to the cosmological constant. 

The model produces non-zero effective cosmological constant
due to the phenomenon of cosmological symmetry breaking. 
We have provided a formalism to compute the curvature scalar $R$ in 
this model. The final result is very simple and given by the 
Eqs. \ref{eq:VacH} and \ref{eq:v}. These relate $R$ to other parameters
in the theory.

The model contains
some parameters $1/\beta$ and $\lambda$ which take very small values. 
We have also assumed that the Weyl meson coupling $f$ is very small.
The parameter $1/\beta$ is related to the gravitational constant and its
smallness is well known. We have shown that quantum corrections to 
$\lambda$ are constrained by local scale invariance. 
In any case this scalar field 
self coupling parameter is expected to depend logarithmically on the
renormalization scale and hence is not likely to suffer from any fine
tuning problems. 
The loop corrections to $f$ are also suppressed by additional powers 
of the coupling $f$ besides the gravitational coupling. Its renormalization
is intrinsically tied to quantum gravity effects since $S_\mu$ does not 
couple directly to Standard model particles. Hence we are unable to properly
address the issue of renormalization group evolution of this coupling in 
the absence of a consistent theory of quantum gravity. 
The effective cosmological constant or equivalently the curvature scalar
is related to other parameters of the theory by the formulas, Eqs. 
\ref{eq:VacH} and \ref{eq:v}. Hence the curvature scalar is not expected
to suffer from any fine tuning problems as long as the remaining 
parameters do not require any fine tuning. If we ignore quantum
gravity effects we expect that this is true in the present model.


\end{spacing}
\begin{spacing}{1}
\begin{small}

\end{small}
\end{spacing}
\end{document}